# 3D Magnetic Analysis of the CMS Magnet

V.I. Klioukhine[1], D. Campi[2], B. Curé[2], A. Desirelli[2], S. Farinon[3], H. Gerwig[2], D. Green[1], J.P. Grillet[2], A. Hervé[2], F. Kircher[4], B. Levesy[4], R. Loveless[5], R.P. Smith[1]

[1] FNAL, Batavia, IL, USA, [2] CERN, Geneva, Switzerland, [3] INFN, Genova, Italy,
[4] CEA/Saclay, Gif-sur-Yvette, France, [5] University of Wisconsin, Madison, WI, USA.

*Abstract*--The CMS magnetic system consists of a superconducting solenoid coil, 12.5 m long and 6 m free bore diameter, and of an iron flux-return yoke, which includes the central barrel, two end-caps and the ferromagnetic parts of the hadronic forward calorimeter. The magnetic flux density in the center of the solenoid is 4 T. To carry out the magnetic analysis of the CMS magnetic system, several 3D models were developed to perform magnetic field and force calculations using the Vector Fields code TOSCA. The analysis includes a study of the general field behavior, the calculation of the forces on the coil generated by small axial, radial displacements and angular tilts, the calculation of the forces on the ferromagnetic parts, the calculation of the fringe field outside the magnetic system, and a study of the field level in the chimneys for the current leads and the cryogenic lines. A procedure to reconstruct the field inside a cylindrical volume starting from the values of the magnetic flux density on the cylinder surface is considered. Special TOSCA-GEANT interface tools have being developed to input the calculated magnetic field into the detector simulation package.

## I. INTRODUCTION

This paper describes the 3D calculations of the magnetic field and forces in the Compact Muon Solenoid (CMS) magnetic system [1], which consists of a superconducting solenoid coil, 12.5 m long and 6 m free bore diameter, and of an iron flux-return yoke.

The yoke, 14 m outer diameter, includes the central barrel surrounding the coil, two end-caps located at each end of the solenoid, and the ferromagnetic parts of the hadronic forward calorimeter arranged downstream of the end-caps. The barrel is subdivided into five rings each comprising three layers joined by connecting brackets. An additional fourth layer, the tail catcher, exists only inside the central barrel ring. Each end-cap consists of one small nose disk and four large disks mounted axially along the solenoid axis line. The design value of the magnetic flux density in the coil center is 4 T.

The calculations were performed with the finite-element analysis code TOSCA [2]. The goal of the 3D analysis was to study the influence of azimuthally asymmetric parts of the system on the field behavior, to calculate the forces caused by the coil misalignments within the iron yoke, to prepare the 3D field map for detector simulations, and to investigate the reconstruction of the field inside the coil volume starting from measurements of the field values near the coil inner surface.

## II. THE MODEL DESCRIPTION

To perform the magnetic analysis of the CMS magnetic system with the TOSCA code, several 3D models were developed [3], [4], [5].

The present model shown in Fig. 1 with one eighth of the magnetic flux return yoke takes into account such azimuthally asymmetric ferromagnetic parts of the system as brackets between the barrel ring layers, the tail catcher, and uses a coil model consisting of five concentric but axially separated modules. The finite-element mesh is constructed from 107784 nodes.

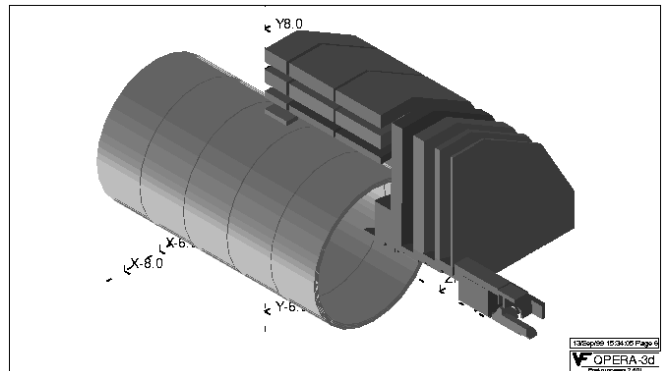

Fig. 1. TOSCA computer model of the CMS magnetic system

To describe the model, a Cartesian coordinate system with the origin placed in the center of the solenoid was used. The direction of the Z-axis is along the solenoid axis. The direction of the Y-axis is upward.

In the model, the superconducting coil modules have length 2443.87 mm and they are spaced axially from one another by 47.6 mm. Each module consists of four coaxial layers of current each of radial thickness 20.63 mm at radii 3174.755, 3239.525, 3304.295 and 3369.065 mm. The overall axial length of the assembly is 12402.39 mm. The locations of the current layers in the calculations correspond to the positions of the superconducting cable in the conductor when the coil is at cryogenic temperature.

The total current in the coil is 42.51 MA-turns [6] which gives 4.08 T in the center of solenoid. The maximum magnetic flux densities, **B**, near the superconducting cable in each layer of the coil are 4.6, 3.5, 2.5 and 1.5 T. The full geometry of the flux return yoke and the steel magnetic properties are described elsewhere [5].

## III. GENERAL FIELD BEHAVIOR

The magnetic flux density contours in the horizontal XOZ-plane are shown in Fig. 2. Contours are plotted every 0.5 T. The maximum value of 5 T is reached inside the hadronic end-cap calorimeter support tube. The line near Z = 2 m corresponds to B = 4 T.

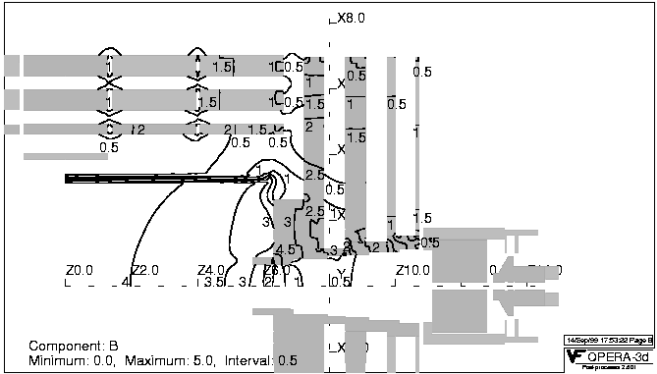

Fig. 2. The CMS magnetic flux density map in horizontal plane

From this map obtained using a special map extraction procedure developed for this purpose [7], one can see that the field near the coil edges, between the barrel and the end-cap, inside the nose and inside and between the first two end-cap disks is strongly inhomogeneous. The magnetic flux density has non-negligible radial component, $B_r$, and large variation of axial component, $B_z$, in the regions of the first layer of the end-cap muon chambers as was shown earlier in [8]. Between the barrel ring layers near the brackets, the field variations are large as well. This complexity of the field behavior requires the use of a 3D field map in detector simulations outside the coil volume.

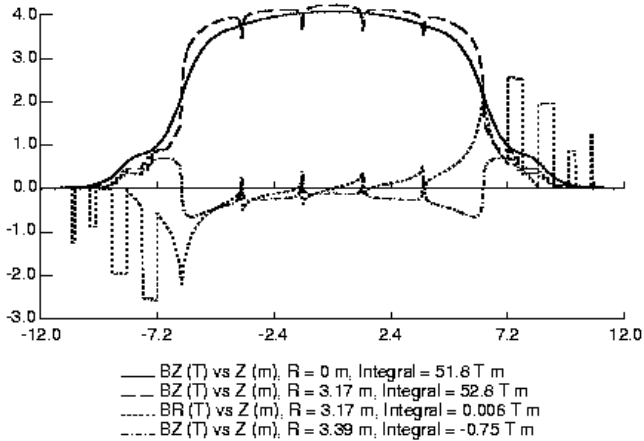

Fig. 3. $B_z$ and $B_r$ at radii 0, 3.17, and 3.39 m from Z-axis

In Fig. 3, $B_r$ and $B_z$ are displayed along the coil axis and near the coil inner and outer radii. This figure shows the influence of the gaps between the coil modules on the field behavior near the coil. At the coil ends the absolute values of the axial and radial components are approximately equal to one half the central field value. Along the line at the coil inner radius the $B_r$ values in the first two end-cap disks are comparable to this. In the last two end-cap disks the radial component is about half that in the first two disks.

The stray field outside the yoke was calculated out to a radius of 50 m from the coil axis. As shown in Fig. 4, at 0.5 m radially outside the yoke the stray field has maxima of 65-188 mT near the gaps between the barrel rings and between the end-cap disks.

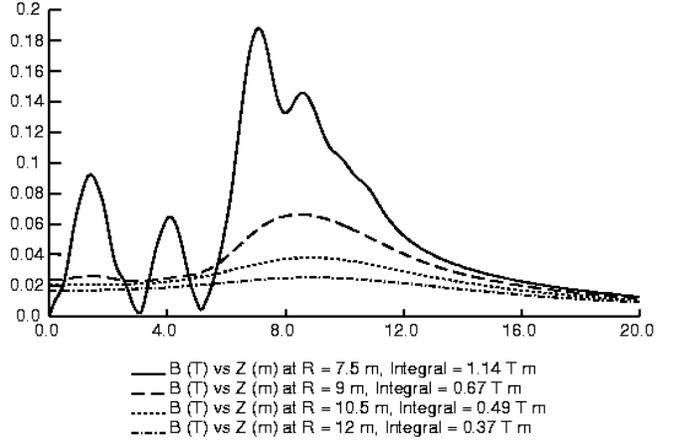

Fig. 4. Stray fields at radii 7.5, 9, 10.5, and 12 m from Z-axis

The stray field in the barrel region decreases faster than in the end-cap regions. At radius of 50 m the value of the fringe field is approximately 0.4 mT.

Two chimneys penetrate the CMS barrel yoke: the vertical chimney contains the cryogenic lines, and the second one, inclined at 30° to the vertical, contains the electrical leads. A special 4π-geometry model was developed to estimate the field value inside these chimneys. In Fig. 5 the behavior of $B_r$ and $B_z$ inside the chimneys, starting from the coil cryostat, are displayed.

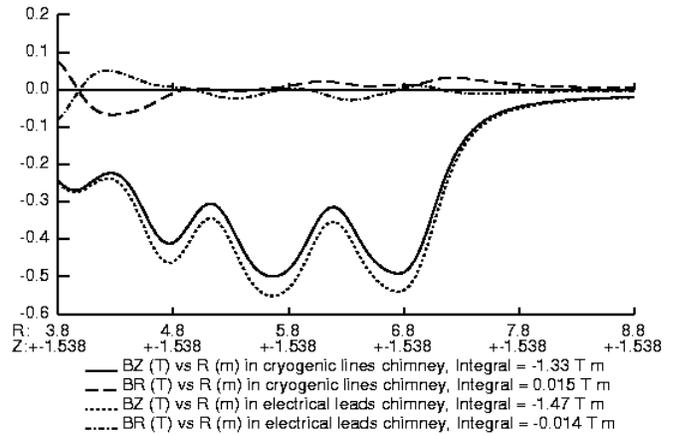

Fig. 5. Axial and radial fields inside the chimneys

The radial component is 50-80 mT between the cryostat and barrel yoke. The axial component oscillates, increasing inside the barrel steel plates and decreasing between them. The maximum $B_z$ absolute value is 0.55 T.

## IV. Forces on Ferro-magnetic Parts

The forces acting on various parts of the yoke were calculated by integrating the Maxwell Stress tensor over the volumes or the surfaces of the yoke pieces [5]. When the integration surfaces pass through ferromagnetic material, an infinitely thin gap at the integration surface is assumed and the following formula is used:

$$\mathbf{F} = \int_S [\mathbf{H}_a(\mathbf{B} \cdot \mathbf{n}) - \mu_0(\mathbf{H}_a \cdot \mathbf{H}_a)\mathbf{n}/2]dS, \quad (1)$$

where $\mathbf{H}_a = \mathbf{H} + ((\mathbf{B}/\mu_0 - \mathbf{H}) \cdot \mathbf{n})\mathbf{n}$, $\mathbf{B}$ is a vector of the magnetic flux density, $\mathbf{H}$ is a vector of the magnetic field strength, $\mathbf{n}$ is a normal unit vector external to surface of region, and $\mu_0 = 4\pi \, 10^{-7}$ (V s)/(A m).

The axial magnetic forces on the first end-cap disk, the nose, and the hadronic end-cap calorimeter support tube are 58.6 MN. Taking into account the inward deflection of the end-cap disks by 30 mm after reaching the nominal current in the coil, this axial force increases to 59.3 MN. These values are close to the values obtained earlier in [9]. The full table of forces on the yoke iron parts is presented in [5].

To get a feeling of the order of magnitude of the forces, an engineering estimation based on a simple version of the previous formula, which considers only the normal components of the fields, was done by S. Vorojtsov [5]. To cross-check the TOSCA results, the simplified POISCR [10] 2D computer model was used. Both of those alternative approaches yielded results comparable with the TOSCA calculations.

## V. Forces on the Coil

In the described model with centered coil, the axial force on the coil module next to the central one is 26 MN and that on the external module is 118 MN.

In the models with decentered coil (145339 nodes) a calculation of the axial force generated by a small coil axial displacement yielded a value of 843 kN/cm, which agrees well with the previous results obtained with the ANSYS 2D [11], CASTEM 2000 [12], and TOSCA [3,4] codes.

From nine models (117611-148311 nodes) used for the calculation of the radial force caused by a small coil radial displacement the average value of the force was 59 ± 19 kN/cm. This value is lower than the value of the radial force obtained in [12] and [3,4]. The force acts in the direction of increasing the coil radial displacement.

From six models (117611-149581 nodes) used for the calculation of torque caused by a small coil angular tilt the average value of the torque was 914 ± 98 kN m/mrad. This value is lower than the value of torque obtained earlier in [12] and higher than the value obtained in [3,4]. The torque direction tends to increase the tilt of the coil. As shown in [3,12] the force and torque dependence from the coil displacements and tilt is linear for the misalignments considered.

## VI. Field Reconstruction

A possibility to determine the field inside a cylindrical volume using the values of $B_z$ on the cylinder surface is investigated. The field boundary values extracted from TOSCA are used for $B_z$ and $B_r$ reconstruction in the volume with help of the CERN codes MAGFIT2 [13] and MAGFIT [14]. Both these programs give approximately the same results but need modification to obtain the desired accuracy.

In the fitting, 10 radial points (spaced 0.3 m, starting from the coil axis), 16 azimuthal points (spaced 22.5°), 113 axial points (spaced 0.1 m), for a total of 2002 points, were used on the surface of a cylinder with a radius of 2.95 m and with a length of 11.2 m. In Fig. 6 the values of $B_z$ on this surface are plotted in contours with an interval 102.5 mT between them. The $B_z$ minimum and maximum values are 2.98 and 4.21 T, respectively.

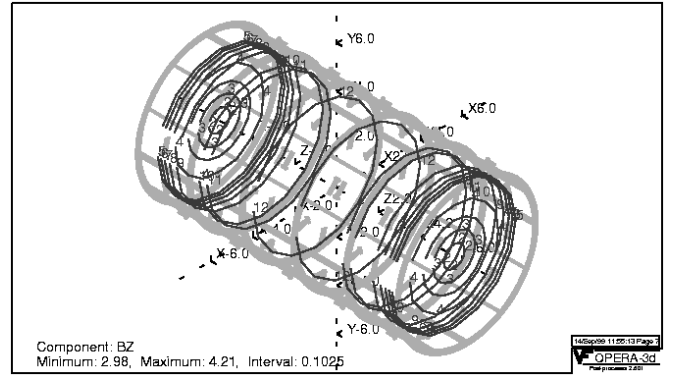

Fig. 6. Axial component of **B** near the cryostat inner surface

In the MAGFIT2 code the flux densities are expanded in Fourier series along the cylinder surface, around rings at the ends of the cylinder surface, and on discs at the ends of the cylinder. The total number of coefficients used in the expansion is 924.

With this parameterization the axial and radial components of the magnetic flux density along the line at radius 2.95 m and at zero azimuth were reconstructed and then the comparisons with the input field values were done.

The accuracy of the $B_z$ reconstruction is 0.75 mT at the edges and 0.25 mT in the middle of the given line, the accuracy of the $B_r$ is 19.5 mT at the edges and around 5 mT in the middle.

The number of radial, azimuthal, and axial points chosen in the fitting were nearly equal to the maxima allowed in the present version of MAGFIT2. It is believed that increasing those limits in the code will enable the program to predict the field more accurately.

## VII. Conclusion

The substantial inhomogeneity of the CMS magnetic field outside the coil volume requires the use of 3D modeling to prepare a field map for detector simulations. The special TOSCA-GEANT interface tools were developed to perform this task.

The simulations of the CMS magnetic system with several TOSCA 3D models give the forces acting on various parts of the yoke in agreement with earlier calculations.

In the TOSCA 3D models the forces acting on the coil due to the coil radial misalignments and tilts are smaller than predicted in previous models. The axial forces due to the coil axial misalignments are in agreement with earlier force calculations.

A possibility to determine the field inside a cylindrical volume by making measurements of axial component of the magnetic flux density on the surface is investigated. The existing software will need modification to predict the field more accurately.